\documentclass[format=acmsmall, review=false, screen=true, nonacm=true]{acmart}
\usepackage{booktabs} 

\renewcommand\footnotetextcopyrightpermission[1]{}
\setcitestyle{authoryear}

\title[Timing on Coupon Effectiveness]{Estimating the Effect of Timing on Coupon Effectiveness}

\author{Deddy Jobson}
\affiliation{%
  \institution{Mercari, Inc.}
  \city{Tokyo}
  \country{Japan}}
\email{deddy@mercari.com}

\begin{abstract}
The coupon incentive is one of the most common tools marketers use to court users to engage with a business at various stages of the customer life cycle. A variety of factors can affect the effectiveness of a coupon incentive on users, timing being one of them. We hypothesize that coupons can be more effective when delivered at critical times in the customer journey, right when a user is engaging with the platform. Verifying such a hypothesis would typically require real time event-triggered coupon distribution software that may be too expensive to implement. In this paper, we propose a framework in which we apply causal inference on "natural randomized control trial experiments" to measure the effectiveness of sending coupons at the right time to users without requiring a dedicated AB test. We demonstrate the usefulness of our framework in the case of a user onboarding coupon campaign held in our company and show how the results can lead to correct data-driven decisions for the business. Furthermore, in order to test the generalizability of our framework, and to make our research more reproducible, we apply our framework on a user retention campaign with a publicly available dataset. 

\end{abstract}

\keywords{e-commerce, causal inference, uplift modeling, coupon timing}

\begin{document}

\maketitle

\renewcommand{\thefootnote}{\fnsymbol{footnote}}
\footnotetext[1]{A version of this paper was published in the \emph{Proceedings of the 1st Workshop on End-End Customer Journey Optimization}, co-located with KDD 2022, August 15, 2022, Washington, DC. This arXiv version is the author's preprint.}
\renewcommand{\thefootnote}{\arabic{footnote}}
\setcounter{footnote}{0}

\section{Introduction}

It goes without saying that marketing campaigns have played a big role in the growth of businesses. A plethora of strategies are often employed in the marketing mix such as advertisements, recommendations, loyalty programs, etc. One of the more effective strategies that is commonly used is the coupon incentive. Coupon incentives can be used to motivate users to overcome hurdles and better engage with the platform. Given the important role of coupons in the marketing mix, and the extent to which businesses rely on them \cite{shaffer_competitive_1995}, it is imperative to get the maximum value out of the coupons distributed. 

There are many ways to go about maximizing the value obtained from incentives. One approach is to tune the value of the incentive (for example, 10\% off vs 30\% off coupons) and trade off two business metrics: the desired return on investment (ROI), and the number of users converted. Another common approach is to consider the heterogeneity of users' response to coupons and find the best segment of users to give incentives. This approach is taken by X-learners \cite{kunzel_metalearners_2019} and uplift modeling \cite{gutierrez_causal_2017}. In uplift modeling, we estimate the conditional average treatment effect \cite{abrevaya_estimating_2015} (the uplift) of each user had they received the coupon; those with a sufficiently high estimated uplift (decided by a threshold based on business requirements) are given incentives. Combining the above two approaches, one can design a framework that performs mathematical optimization on the predictions made by the uplift model to find the best value of coupon to give to each user \cite{shen_framework_2021}.

One more way to improve the effectiveness of coupons, and this will be the focus of our paper, is to optimize on the timing of coupon distribution. One can use statistical models to estimate the best time to deliver coupons for each user \cite{johnson_whom_2013}. A simpler heuristic that can be used to improve the timing of the distribution of coupons is to deliver coupons right at the moment when users are actively engaging with the service. In the case of onboarding campaigns, the coupon incentives may be more effective when coupons are given right after registration. In the case of retention or churn campaigns, we can try to give coupons closer to when the user interacted with the platform. In general, users can be hypothesized to not have as much of a hurdle to making a conversion at the moment when they are engaging with the platform. Distributing coupons exactly when users are active would require the development of a real time coupon distribution service, one that can be triggered by customer events (such as app open, item view, etc.). 

Implementing such coupon distribution services would incur considerable engineering effort and may not be undertaken without substantial evidence of the possible benefits. While some studies have shown the importance of time when designing coupons \cite{tellis_tackling_1995}, we cannot say for sure to what extent the results are extendable to one's particular business context. A randomized control trial (RCT) is the best way to be confident that such a coupon distribution service will be useful. But such RCTs cannot be performed without the prior implementation of coupon distribution services themselves thus resulting in a catch-22. 

In this paper, we explain how we could get around the catch-22 in our company, a large e-commerce platform with millions of monthly active users. We do this by applying causal inference \cite{pearl_causal_2009} methods on a preexisting RCT conducted for another purpose, successfully demonstrating the value of distributing coupons for the purpose of user onboarding as quickly as possible after they have completed their account registration.

The following is the summary of our contributions:
\begin{itemize}
    \item We highlight the importance of timing when distributing coupon incentives.
    \item We propose a framework to confirm the benefit of the timing of coupon distribution without the need for a dedicated AB test (and the expensive real time coupon distribution service that would be required to perform such a test). 
    \item We demonstrate the value of our framework in the case of user onboarding in our company by measuring the effect of sending coupons earlier in the user journey.  
    \item We further demonstrate the generalizability of our framework by measuring the effect of sending coupons earlier to churn users for the purpose of user retention.
\end{itemize}

The rest of the paper is structured as follows. In Section \ref{experiment_setup}, we state the estimation problem we face using the language of causal inference, introduce our framework, and discuss alternative methods. We then discuss our findings from the onboarding campaign in Section \ref{results}. To further demonstrate the effectiveness of our method, we additionally apply our framework to analyze the effectiveness of timing in a user retention campaign in Section \ref{reproducible_experiment}. We use a public dataset in the case of retention to make our research easier to reproduce. Finally, we mention some limitations of our method in Section \ref{limitations} and conclude in Section \ref{conclusion}.

\section{User Onboarding}
\label{experiment_setup}

\subsection{Experiment Setup}

In the past, our company did not possess event-triggered real time coupon distribution services and instead relied on hourly batch cron jobs to deliver incentives to users. At the turn of each hour, all users who registered in the previous hour were gathered and randomly allocated to the control and test groups.
The randomized allocation into control and test groups was done for the purpose of evaluating the effectiveness of the campaign overall.
We can make use of this RCT data to measure the effectiveness of timing on coupon effectiveness.
Note that this RCT was originally created for the purpose of evaluating the effectiveness of the campaign in general and not the timing of the coupon which is what we are interested in. 
This kind of data is prevalent in our company and we expect the same to be true in the case of other companies as well thus making our method generally usable by other companies too.
Not all test group users were given coupons; users in the test group were only given coupons if they had not yet made a purchase by the time of coupon distribution. This was done to save on costs; ignoring this fact would have led to a biased estimation of the importance of the timing of coupon distribution, as we will show when we discuss the causal diagram. Last but not the least, we do not distribute coupons at night so as to avoid disturbing the user. While the user selection is randomly chosen by the marketer, the time since registration when the coupon is distributed is out of the marketer's control giving it some properties of a natural experiment \cite{leatherdale_natural_2019}. It will therefore be better to call the above RCT a natural randomized control trial experiment for our purposes.

\subsection{Target Users}
We gather a subset of users who created an account over a period of one month and check whether or not they made a purchase within a certain number of hours of creating an account. We make use of the following covariates:

\begin{itemize}
\item hour of the day of registration (H)
\item minute of the hour of registration (M)
\end{itemize}

We preprocess the features by standardization to make the estimated coefficients unitless. While different methods of feature engineering exist to account for the cyclical nature of time-related features, we do not need them here. This is because we don't distribute coupons at night, and users who register at the same minute of the hour get coupons with the same delay irrespective of the hour of the day they receive coupons. The target users are therefore not affected by the arbitrary discontinuity that normally gets imposed on cyclic time-related features like the hour of the day around midnight or the minute of the hour around the turn of the hour. 

\begin{figure}
  \centering
  \includegraphics[width=0.9\linewidth]{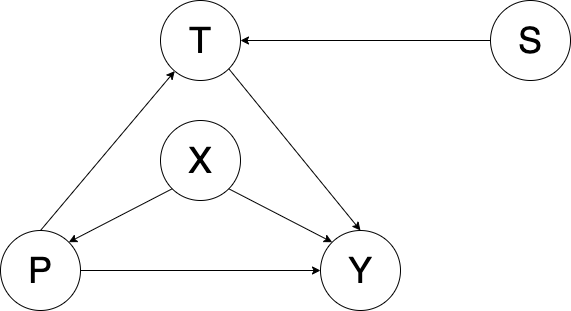}
  \caption{The causal model for our onboarding campaign}
  \label{fig:causal_diagram}
\end{figure}

We make use of a causal model depicted by the Directed Acyclic Graph (DAG) in Figure \ref{fig:causal_diagram}. The following is the terminology of our DAG:
\begin{itemize}
\item S - segment of the user which is decided at the time of account creation.
\item Y - dependant variable; in this case, the probability of a future purchase.
\item T - whether or not the user was actually given coupons.
\item P - whether or not the user made a purchase prior to the coupon distribution schedule. 
\item X - user attributes like hour of user registration and minute of user registration. 
\end{itemize}

The two covariates we use, hour of user registration and minute of user registration, will be represented by H and M respectively. 
$$H,M \in X$$

From the causal graph, we can see that by controlling P, we can estimate the effect of treatment (T) on the dependant variable (Y) without bias. However, our goal is to get an estimate for the interaction between treatment (T) and one of the covariates X: the minute of the hour of registration (M). To estimate the interaction correctly, we need to measure the joint causal effect of both the interacting variables of interest on the target variable. In the language of Pearl \cite{pearl_transportability_2011} we want to estimate $P(Y|do(T),do(M))$ without bias. While controlling for P will block the backdoor path for the variable T, it will also block one of the forward paths between X and Y. Therefore, we cannot directly get an unbiased estimate for the interaction between M and T towards Y. 

We can, however, get an unbiased estimate for the interaction between X and S, the segment of the user. In this case, we should simply not condition on P in order to get the total joint causal effect of M and S on Y $(P(Y|do(S),do(M)))$. If we do condition on P or in other words, if we consider only users who did not make a purchase as of the time of coupon distribution, then we would arrive at biased results which would lead to incorrect business decisions down the road. The causal diagram thus helps us prevent making common mistakes such as controlling for all pretreatment variables \cite{greenland_causal_1999} as we may have had naively done.

\begin{figure}
  \centering
  \includegraphics[width=0.9\linewidth]{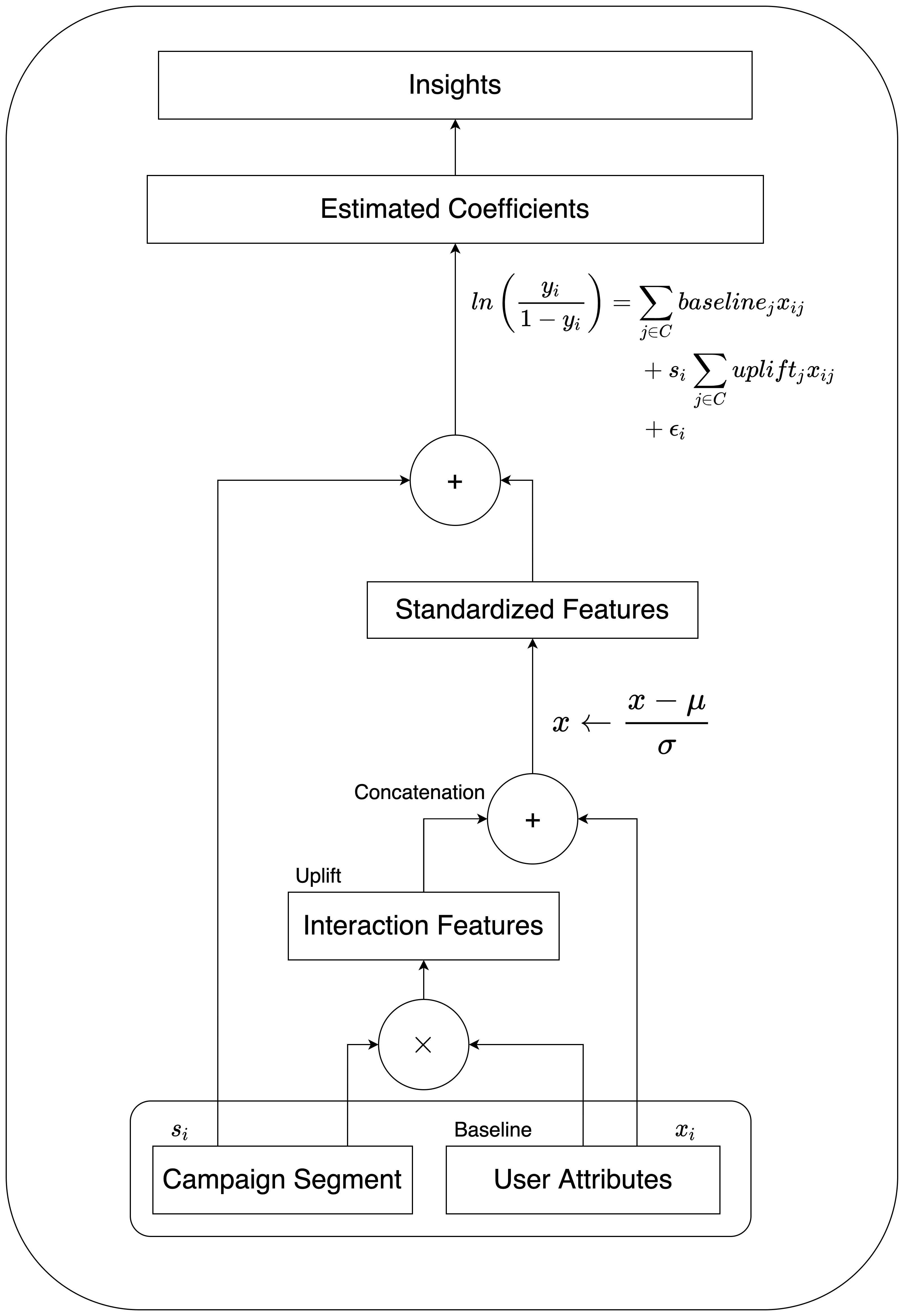}
  \caption{Our framework for estimating the effectiveness of timing on coupons. The time-related feature of interest is one of the user attributes.}
  \label{fig:analytics_schema}
\end{figure}

\subsection{Related Methods}
As far as we are aware, this is the first study to make use of do calculus \cite{pearl_transportability_2011} from causal inference to measure the unbiased effect of timing on the effectiveness of coupons. However, one can make use of other ways to measure the interaction between timing and treatment variable to achieve similar outcomes. 

The problem of estimating interaction has especially garnered attention from the perspective of clinical trials. It is of great interest to the research community which kinds of people respond well to treatment for a disease. The problem is not too dissimilar to ours. However, the methodology used is quite different. While we use a logistic regressor with interaction terms, they measure the treatment effect for various sub groups \cite{brankovic_understanding_2019} of the target users and look for differences that can be explained by changes in covariates. A limitation of doing so is that the definition of the sub groups, if optimized on, can lead to an inflation of the type-1 error of the estimate of the interaction \cite{altman_dangers_1994}. Other methods have been proposed to overcome the limitation while still capturing non-linear interactions \cite{royston_new_2004,bonetti_graphical_2000,sauerbrei_detecting_2007}.

In the domain of marketing, interactions between covariates and treatment have been approached under the name of conjoint analysis \cite{klenosky_assessing_1996,uchida_demand_2014}. These methods typically require dedicated AB tests and therefore can't be used as is in our scenario. Our method doesn't require a dedicated AB test and can instead be applied on data obtained from preexisting AB tests commonly conducted to measure the business impact of coupon campaigns.

In order to measure the interaction effect between M and T, an alternative approach would be to treat S as an instrumental variable \cite{nogueira_methods_2022} and then estimate $P(Y|do(T),do(M))$. This can be done since S is independent of any confounders and can only affect Y through T. In our case, however, the interaction between M and S would give us the effect of coupon timing (from the perspective of the user) on the target users as a whole and will suffice to convince the stakeholders of the necessity (or lack, depending on the results) of a real time coupon distribution service.

\begin{table*}
  \caption{The coefficients of the uplift model for user onboarding}
  \label{tab:coeffs}
  \begin{tabular}{|c|c|c|c|c|c|c|}
    \toprule
    coefficient  &  mean   & std err & t   & P>|t|   & [0.025  & 0.975] \\
    \midrule
    baseline\_intercept &	-2.6104 &	0.16 &	-16.28 &	0.00 &	-2.925 &	-2.296 \\
    \hline
    baseline\_reg\_hour &	0.0468 &	0.028 &	1.67 &	0.09 &	-0.008 &	0.102 \\
    \hline
    baseline\_reg\_minute &	-0.0295 &	0.028 &	-1.06 &	0.29 &	-0.084 &	0.025 \\
    \hline
    uplift\_intercept &	0.1894 &	0.178 &	1.06 &	0.29 &	-0.159 &	0.538 \\
    \hline
    uplift\_reg\_hour &	-0.0198 &	0.057 &	-0.35 &	0.73 &	-0.132 &	0.092 \\
    \hline
    uplift\_reg\_minute &	0.0617 &	0.031 &	1.96 &	0.05 &	0.0 &.123 \\
  \bottomrule
\end{tabular}
\end{table*}

\subsection{Uplift Modeling}
Once we use a causal graph to find which covariates to condition on, we can use an uplift model to estimate the interaction of user registration time and treatment on the probability of future purchase which we will interchangeably referred to as the Buyer Conversion Rate (BCR). While there are multiple approaches to uplift modeling like the two-model approach \cite{radcliffe_using_2007} and target transformation \cite{jaskowski_uplift_2012} and multiple algorithms to use for estimation like decision trees and neural networks, we find that adding interaction terms (new features created by multiplying the treatment variable and each covariate) on a logistic regression model gives us the most useful uplift model for the purpose of obtaining the most interpretable insights.  

The following is our formulation. For user i and covariates $C$, we have 
$$ln\left(\frac{y_i}{1-y_i}\right) = \sum_{j\in C} baseline_j x_{ij} + s_i  \sum_{j\in C} uplift_j x_{ij} + \epsilon_i$$
where each $x_i$ is a vector of covariates (including a bias term), \textbf{$baseline$} and \textbf{$uplift$} are the weight vectors with as many terms as covariates, $s_i$ is an indicator variable corresponding to the segment of the user (1 for treatment and 0 for control), and $\epsilon_i$ is the residual of the prediction made for each user.

Our entire framework is depicted in Figure \ref{fig:analytics_schema}. As can be seen in the figure, we first create interaction terms between the covariates of interest and treatment. We then standardize all the features and train the uplift model. Finally, we obtain the insights regarding the effect of distribution timing on the treatment effect of the coupon from the coefficients estimated by the uplift model.

\section{Results}
\label{results}

In table \ref{tab:coeffs}, we show the mean and confidence intervals of all the coefficients of the linear component of the uplift model. The coefficients marked baseline\_* indicate the baseline coefficients for users in both the control group and treatment group while those marked uplift\_* indicate the incremental effect on the log odds of making a purchase caused by treatment on the users of the treatment group. We find that the uplift coefficient for registration minute is positive and statistically significant while the same is not true for the uplift coefficient of the registration hour. Furthermore, we see that the effect of the baseline hour is stronger than that of the baseline minute while the opposite is true in the case of the uplift coefficients. From the above, we infer that the time between user registration and coupon distribution plays an important role in the effectiveness of user onboarding coupons. With this new found evidence, we can now implement a real time coupon distribution service to perform more AB tests to measure the overall business impact achieved by delivering coupons right after users create their account.

We can also create counterfactual plots with the predictions of the uplift model as done in Figure \ref{fig:counterfactual_prediction_onboarding} to visualize the benefit of distributing coupons to users right after they register. We see that control group users who register later in the hour have a slightly lower BCR; we don't expect that trend to change by delivering coupons. The positive effect on the treatment users, however, can be seen quite clearly seen, thus confirming what we learned from the uplift coefficients.

\begin{figure}
  \centering
  \includegraphics[width=0.9\linewidth]{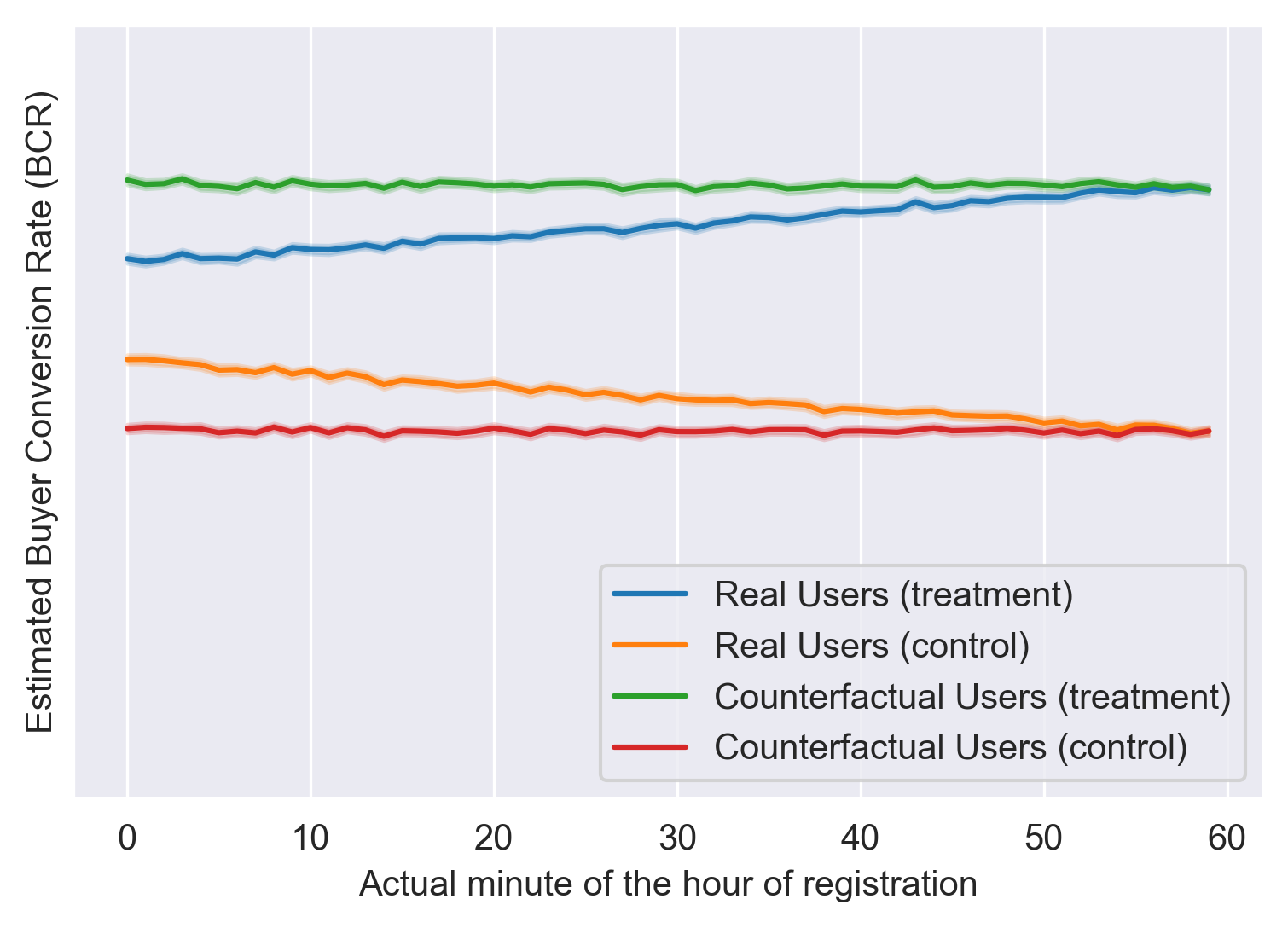}
  \caption{Comparison of actual and counterfactual BCR of treatment and control users for different registration times for user onboarding. The counterfactual case is where users register at the turn of the hour thus receiving coupons immediately upon registration.}
  \label{fig:counterfactual_prediction_onboarding}
\end{figure}

A limitation of the above experiment is that the data we worked with is proprietary and so the experiment cannot be easily verified by other researchers. To make our work more reproducible, we look for publicly available datasets with AB test data that also contain time-related features in order to conduct more reproducible experiments.

\section{User Retention}
\label{reproducible_experiment}

While we could not find a public dataset with AB test data from onboarding campaigns to exactly replicate the previous experiment, we did find a public dataset \cite{dataset_x5_2019} that can help us evaluate the effectiveness of the timing of coupons in another stage of the customer journey: retention. We shall first state the problem.
\subsection{Data}
Users who have just made a purchase can be considered to be in "the zone" \cite{young_zone_1999} for making purchases and one can hypothesize that targeting users immediately after they make a purchase can better incite more purchases thus getting one more value out of coupons. To test the hypothesis, we need to estimate the interaction between the treatment variable and the time elapsed since the latest purchase.

Lenta is a large Russia hypermarket chain that uses various communication channels in its marketing strategy with its customers. One of them is targeted advertising campaigns for regular consumer goods. They have made publicly available the results of an AB test they performed on users and along with that, they have helpfully provided the purchase logs of the target users. Using the data provided, we can treat the campaign to be a retention campaign (one which attempts to reduce the inter purchase time) and estimate the interaction between treatment and the time elapsed since the last purchase. One benefit in retention as opposed to the earlier example is that we can make use of other user attributes like gender and age data which we could not in the case of user onboarding since such data would not have been available at the time of user account registration. In total, we use the following user attributes from the dataset:
\begin{itemize}
\item gender
\item age
\item purchase recency 
\item purchase frequency
\item the time the user received their first coupon (in the past)
\item the time taken to respond to their first coupon
\end{itemize}

We construct the purchase recency feature by taking the negative of the log of the time difference between the each user's latest purchase and the latest purchase of all users.
After performing the log transformation, we standardize all the numeric features like before using the mean and standard deviation of each feature. 


To deal with missing data, we perform a complete case analysis, thus restricting our analysis (and takeaway insights) to users with no missing data. The only exempted features from deletion is gender for which we simply remove the dummy variable for undefined gender and incorporate it into the intercept.

\begin{table*}
  \caption{The Coefficients of the uplift model on Lenta's dataset}
  \label{tab:coeffs_lenta}
  \begin{tabular}{|c|c|c|c|c|c|c|}
    \toprule
    coefficient  &  mean   & std err & t   & P>|t|   & [0.025  & 0.975] \\
    \midrule
    baseline\_intercept & -0.0196 & 0.116 & -0.168 & 0.867 & -0.248 & 0.209 \\
    \hline
    baseline\_gender\_M & 0.0431 & 0.020 & 2.130 & 0.033 & 0.003 & 0.083 \\
    \hline
    baseline\_gender\_F & 0.0474 & 0.016 & 3.021 & 0.003 & 0.017 & 0.078 \\
    \hline
    baseline\_age & 0.0562 & 0.013 & 4.242 & 0.000 & 0.030 & 0.082 \\
    \hline
    baseline\_latest\_purchase\_time & 0.5282 & 0.008 & 63.349 & 0.000 & 0.512 & 0.545 \\
    \hline
    baseline\_first\_issue\_time & 0.0569 & 0.007 & 7.882 & 0.000 & 0.043 & 0.071 \\
    \hline
    baseline\_issue\_redeem\_delay & 0.1483 & 0.007 & 20.941 & 0.000 & 0.134 & 0.162 \\
    \hline
    baseline\_purchase\_frequency & 0.7635 & 0.011 & 66.631 & 0.000 & 0.741 & 0.786 \\
    \hline
    uplift\_intercept & 1.2551 & 0.233 & 5.393 & 0.000 & 0.799 & 1.711 \\
    \hline
    uplift\_gender\_M & 0.0203 & 0.029 & 0.706 & 0.480 & -0.036 & 0.077 \\
    \hline
    uplift\_gender\_F & 0.0695 & 0.022 & 3.109 & 0.002 & 0.026 & 0.113 \\
    \hline
    uplift\_age & 0.0611 & 0.016 & 3.806 & 0.000 & 0.030 & 0.093 \\
    \hline
    uplift\_latest\_purchase\_time & 0.0202 & 0.061 & 0.329 & 0.742 & -0.100 & 0.141 \\
    \hline
    uplift\_first\_issue\_time & 0.5462 & 0.103 & 5.320 & 0.000 & 0.345 & 0.747 \\
    \hline
    uplift\_issue\_redeem\_delay & 0.0448 & 0.007 & 6.207 & 0.000 & 0.031 & 0.059 \\
    \hline
    uplift\_purchase\_frequency & -0.0559 & 0.015 & -3.833 & 0.000 & -0.085 & -0.027 \\
  \bottomrule
\end{tabular}
\end{table*}

\begin{figure}
  \centering
  \includegraphics[width=0.5\linewidth]{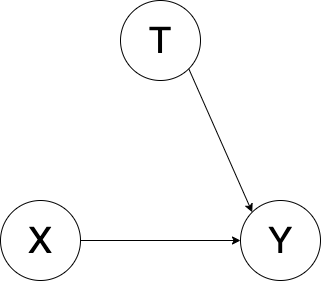}
  \caption{The causal model for Lenta's retention campaign}
  \label{fig:lenta_causal_diagram}
\end{figure}

The causal diagram for this scenario depicted by Figure \ref{fig:lenta_causal_diagram} is considerably simpler than in the case of the user onboarding example. The following is our terminology:
\begin{itemize}
\item X - user attributes like gender, age, purchase frequency, etc.
\item Y - dependant variable; whether or not the user made a subsequent purchase
\item T - whether or not the user was given coupons
\end{itemize}

Our objective is to measure the interaction between the time elapsed since the latest purchase and treatment. We use the same framework as in the case of user onboarding (Figure \ref{fig:analytics_schema}) for user retention. We make our source code available on GitHub\footnote{https://github.com/deddyjobson/Estimating-the-Effect-of-Timing-on-Coupon-Effectiveness}.

\subsection{Results}

In table \ref{tab:coeffs_lenta}, we show the obtained estimates from the model. We see that the coefficient for "latest\_purchase\_time\_baseline" is strongly positive even when controlling for purchase count. This corroborates with our hypothesis that users with more recent purchases could be in "the zone" \cite{young_zone_1999} for shopping around the time of their purchases. We also note that the uplift coefficient for registration minute is weakly positive but not statistically significant. We therefore infer that the gains to be made from implementing an event-triggered coupon distribution service that can deliver coupons to users right after they make a purchase is small.

Looking at the counterfactual predictions of the uplift model can give us a clearer picture of how much of a gain can be achieved by giving coupons right after the user's latest purchase. From Figure \ref{fig:counterfactual_prediction_onboarding}, we don't see much of a difference in the uplift of users across different values of purchase recency, even though the baseline probability changes dramatically. The graph therefore corroborates with what we learned from the coefficients, that the benefit of sending coupons quicker after a user's latest purchase doesn't yield a considerable business impact.  

\begin{figure}
  \centering
  \includegraphics[width=0.9\linewidth]{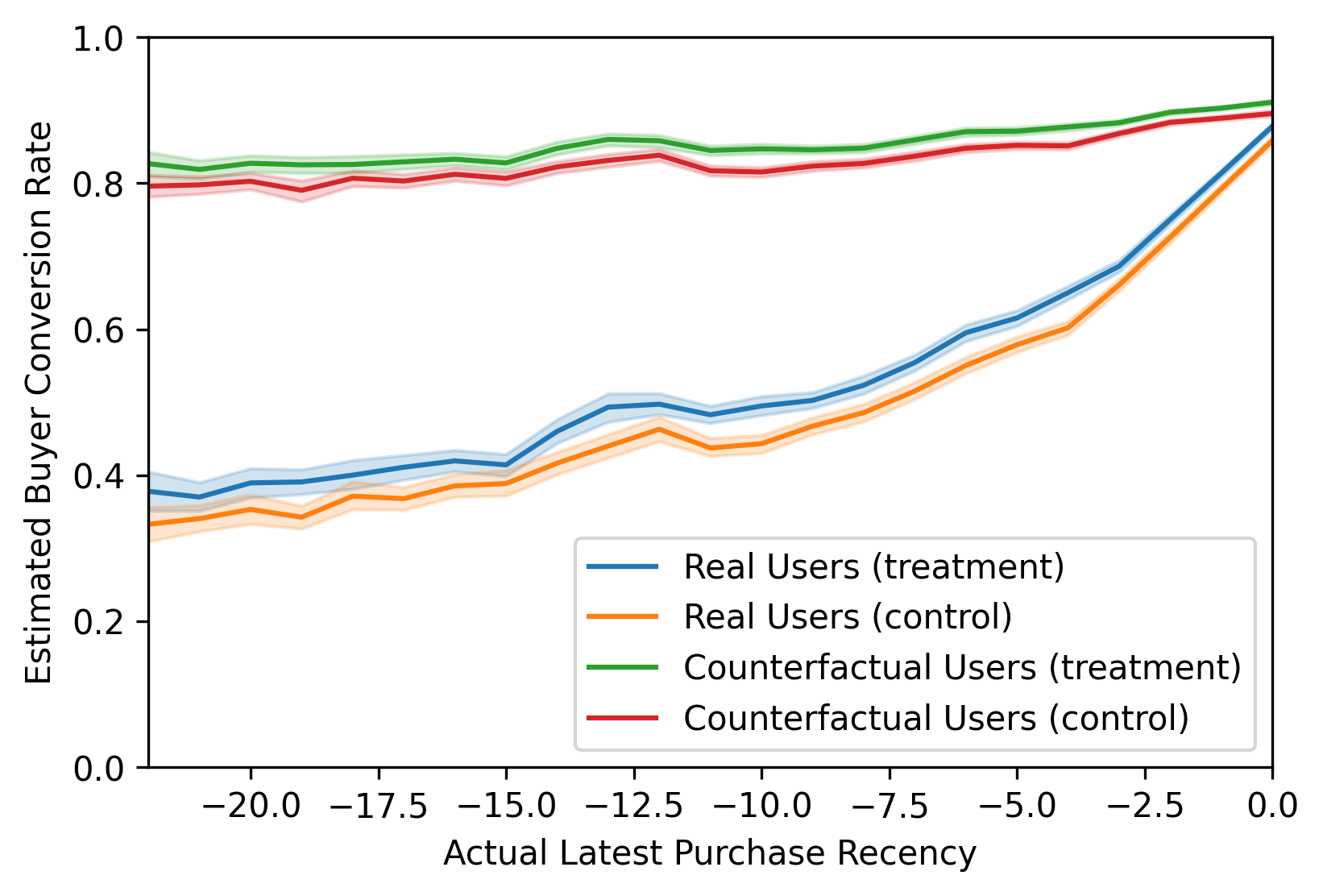}
  \caption{Comparison of actual and counterfactual BCR of treatment and control users for different purchase recency values in a user retention campaign. The counterfactual case is where users receive coupons right after making their latest purchase.}
  \label{fig:counterfactual_prediction_onboarding}
\end{figure}

\section{Limitations}
\label{limitations}
Estimations made through causal diagrams are unbiased conditional on the validity of the causal diagram. The limitations of our research are therefore tied to the limitations of the causal model we used.

In the user onboarding example, we assume the non-existence of unobservable confounders \cite{damour_multi-cause_2019} like income, hobbies, etc. that can arguably have a direct causal effect on both the minute of registration and the BCR. However, we expect that controlling for the hour of registration when estimating the interaction effect of the minute of registration should mitigate the confounding effect induced by these unobservable confounders. 

Fortunately, in the case of the public dataset, we had many covariates at our disposal to shield the effect of unobservable confounders on the covariate of interest (purchase time). In general, we would recommend one to use as many covariates as available that can reduce the bias induced by unobservable confounders.

\section{Conclusion}
\label{conclusion}

In this paper, we propose a framework to confirm the benefit of timing the coupon incentives distributed to users without the need of a dedicated randomized control trial. Our method applies causal inference methods on natural randomized control trial experiments to estimate the interaction between coupon distribution time and coupon effectiveness. We test our framework on a user onboarding and user retention campaign and find that in both cases (to varying extents), the earlier the coupons are given to users after their account creation/latest purchase, the more effective the coupons.



\bibliographystyle{ACM-Reference-Format}
\bibliography{references}

\end{document}